\documentclass[letterpaper, a4paper]{amsart}
\usepackage{amsmath}
\usepackage{latexsym}
\usepackage[all]{xy}
\usepackage{color}
\newtheorem{teorema}{Theorem}[section]
\newtheorem{definicion}[teorema]{Definition}
\include{amslatex}

\usepackage[margin=1.4in]{geometry}
\newtheorem{comentario}[teorema]{Remark}

\numberwithin{equation}{section}
\begin{document}
\begin{title}[On the Unruh effect in spacetimes of maximal proper acceleration]
{On the Unruh effect for hyperbolic observers in spacetimes of maximal proper acceleration}
\end{title}
\maketitle
\begin{center}
\author{Ricardo Gallego Torrom\'e\footnote{Email: rigato39@gmail.com}}
\end{center}

\begin{center}
\address{Department of Mathematics\\
Faculty of Mathematics, Natural Sciences and Information Technologies\\
University of Primorska, Koper, Slovenia}
\end{center}

\begin{abstract}
In this work, the notion of spacetime of maximal proper acceleration is motivated as a weak form to implement general covariance and a generalized form of Einstein's equivalence principle from a physical point of view and the fundamental geometric and kinematic properties of such spaces are discussed. Thereafter the Unruh temperature formula is generalized to the case of hyperbolic observers in spacetimes of maximal proper acceleration. Such a generalization implies the existence of a maximal value for the Unruh temperature. We discuss this result for an electrodynamic model of point charged particles in a spacetime of maximal proper acceleration. It is shown that according to the model, the maximal Unruh temperature must be of order $4.8\cdot 10^{10}/N$ K for current high acceleration electron laser-plasma acceleration systems, where $N$ is the population of the typical bunch.
\end{abstract}
\bigskip


\section{Introduction}
The conjecture on the existence of a maximal proper acceleration has been a matter of analysis and research in different contexts for a long time. Starting as a consequence of Caianiello's {\it quantum geometry} \cite{Caianiello}, maximal acceleration  has been a recurrent topic in field theory \cite{Brandt1983, Brandt 1989}, string field theory \cite{FrolovSanchez 1991,Parentani Potting}, loop quantum gravity \cite{RovelliVidotto}, non-commutative spacetimes \cite{Harikumar et al. 2020,Harikumar et al. 2021,Harikumar et al. 2022}, rainbow gravity \cite{Bhagya et al.}, emergent quantum mechanics \cite{Ricardo05b,Ricardo06,Ricardo2014}, Born's reciprocity and extended versions of the theory of relativity \cite{Schuller,Castro-Perelman 2022}, classical dynamics \cite{Toller2003,Toller2006} and classical electrodynamics \cite{BornInfeld,Caldirola,Ricardo2012,Ricardo2015,Ricardo 2017,Ricardo 2024}, just to mention several frameworks where the notion of maximal proper acceleration makes its appearance. Different frameworks for theories of maximal acceleration have been reviewed in \cite{Ricardo Nicolini 2018}.

From a classical viewpoint, that is, when the word ''acceleration" refers to the change rate with time of the instantaneous velocity vector at a point of a world line, there are arguments supporting the possibility of dynamical systems with an uniform bound in the acceleration, a bound not depending upon initial conditions of motion or particular values of the energy of the system. In particular, attempts to understand the clock hypothesis in the theory of relativity lead naturally to assume that the proper time functional must depend upon acceleration in situations of strong back-reaction in field-particle interaction systems \cite{Mashhoon1990}. The most simple terms in which such dependence can enter in the metric structure is in the form of a linear correction to a Lorentzian metric in the acceleration squared. On the other hand, the notion of maximal proper acceleration is an invariant concept, a property that renders it a natural notion to form the basis for  consistent generalizations of Lorentzian spacetime geometries in a classical setting. Indeed, the class of spacetime models that we will discuss in this paper, described first in \cite{Ricardo2007} and in more detail in  \cite{Ricardo2015} and in \cite{Ricardo 2024}, are general covariant models.

The kinematical aspects of spacetime metrics with maximal proper acceleration have been discussed in detail in \cite{Ricardo2015,Ricardo 2024}. However, in the first part of the present work, a renewed introduction of the concept of spacetime of maximal proper acceleration that we find physically appealing is presented. In our new approach, the underlying Lorentzian structure is vindicated, by invoking the principle of general covariance and the weak equivalence principle, as the geometry that a test particle without acceleration will experience. The metric of maximal proper acceleration builds on this construction as the most simple high order jet metric structure violating the clock hypothesis but  maintaining the causal structure of the underlying Lorentzian structure and the rest of the regularity conditions except for certain world lines where the maximal proper acceleration is formally reached.

In the second part of this work, we consider certain thermal aspects of spacetimes of maximal proper acceleration, concentrating our attention on the Fulling-Davies-Unruh effect. The Unruh temperature formula is not only a consequence of the theory of quantum fields in arbitrary coordinate systems \cite{Fulling,Davies,Unruh 1976}, but together with a form of the equivalence principle, constitutes a fundamental assumption in several approaches to emergent gravity, namely the thermodynamic interpretation of gravity of Jacobson \cite{JacobsonI} and Verlinde's theory of entropic gravity \cite{Verlinde2011}.
Given this pivotal role of the Unruh temperature formula, it is natural to wonder about  its status in theories with maximal proper acceleration.
In this paper we show that the relation between temperature and acceleration described by Unruh temperature can be extended to hyperbolic observers in spacetimes of maximal proper acceleration in a rather direct way. As a consequence it is shown that maximal proper acceleration in spacetimes of maximal proper acceleration \cite{Ricardo2015,Ricardo 2024} leads to a maximal Unruh temperature. The relation between maximal proper acceleration and maximal temperature by means of an extension of Unruh temperature formula has been previously considered in Caianiello's quantum geometry with ambiguous claims \cite{CaianielloLandi, Benedetto Feoli 2015}, in Brandt's theory \cite{Brandt1983}, in Born's reciprocity relativity theory \cite{Castro-Perelman 2024}, in rainbow gravity and its relation with maximal proper acceleration framework \cite{Bhagya et al.} and in the context of Hawking effect in spacetimes of maximal proper acceleration \cite{Ricardo 2022c}, just to mention several relevant investigations. However, by considering the problem from the perspective of spacetimes of maximal proper acceleration and high order jet electrodynamics \cite{Ricardo2012, Ricardo 2017}, a rigourous extension of Unruh formula and a discussion of its extension is provided.

The general theory of spacetimes of maximal proper acceleration does not fix the value of maximal acceleration. However, in situations when the predominant interaction is of electrodynamic nature, we can apply the theory of high order jet electrodynamics as developed in \cite{Ricardo2012,Ricardo 2017} and more recently discussed in \cite{Ricardo 2024}. Such a theory provides a specific value for the maximal proper acceleration of a classical point charged particle being accelerated by electromagnetic means \cite{Ricardo 2017}. Under certain assumptions relative to the stability of the bunches, high order jet electrodynamics implies specific phenomenological consequences potentially testable with current or near future facilities \cite{Ricardo 2019, Ricardo 2024}. In particular, the scale of the maximal proper acceleration decreases with the inverse of the total charge of the system being accelerated, that for bunches of point charged particles, leads to an effective reduction in the value of the maximal proper acceleration until limits where current or near future accelerator facilities can be sensitive.
In this context of spacetimes of maximal proper acceleration, the potential consequences of maximal proper acceleration for the Unruh temperature perceived by a hyperbolic observer are investigated. The standard Unruh temperature formula can be extended to spacetimes with a maximal proper acceleration. It is also shown that in our framework, maximal proper acceleration implies a maximal Unruh temperature. For the theory of high order jet electrodynamics, this leads to the prediction that for current and near future laser-plasma accelerator facilities, there must be a maximal Unruh temperature of order $10^2\, K$.

The structure of the present paper is the following. In {\it section} \ref{Spacetimes of maximal proper acceleration} we discuss the concept of spacetimes of maximal proper acceleration.  First, the use of metric structures depending on the acceleration of test particles is motivated and the notion of {\it metric of maximal proper acceleration} introduced. The argument followed here links the construction of the metric of maximal proper acceleration to the principles of general covariance and a form of Einstein's equivalence principle. After this, the notions of proper time and angle are introduced and compared with the relativistic ones. These constructions illustrate that the fundamental notions of pseudo-Riemannian geometry can be transported to the new setting in a natural way. We continue proving that in a spacetime of maximal proper acceleration of a timelike world line must be bounded. The case that will be mainly considered in this paper is a model for bunches of particles based in high order jet electrodynamics. It is shown that such a model contains a proper acceleration upper bound, which is identified with the value of the maximal proper acceleration. For comparison, other examples of theories with maximal proper acceleration are briefly discussed. Then  the notions of energy and momentum are discussed and the on-shell condition for point particles in spacetimes of maximal proper acceleration derived, taking special attention to several notions of mass that appear in the theory. In {\it section} \ref{On the Unruh effect in spacetimes of maximal proper acceleration}, the requirements for establishing the Unruh temperature formula in standard flat spacetimes (Minkowski spacetime) are discussed. We show that such requirements also hold for hyperbolic observers in spacetimes of maximal proper acceleration and hence, Unruh temperature must hold good. In {\it section} \ref{The Unruh temperature in laser-plasma acceleration for high order jet classical electrodynamics}, as a consequence of the maximal proper acceleration and Unruh temperature formula, the existence of the maximal Unruh temperature is established. It is shown that for the model of classical point electrodynamics discussed in {\it section} \ref{Spacetimes of maximal proper acceleration}, the value of maximal acceleration leads to a maximal Unruh temperature that decreases with the particle population of the bunch.  This particular result has experimental consequences for laser plasma acceleration that, although small, are testable. In {\it section} \ref{Discussion}, a discussion of our theory and results is presented, comparing them with different treatments of the Fulling-Davies-Unruh effect in the context of theories with maximal proper acceleration.
 \section{Spacetimes of maximal proper acceleration} \label{Spacetimes of maximal proper acceleration}
\subsection{Motivation of the notion of spacetimes of maximal proper acceleration}
There are well known quantum mechanical arguments suggesting the existence of a maximal proper acceleration in several dynamical settings \cite{Caianiello,RovelliVidotto,FrolovSanchez 1991,Parentani Potting}. On the other hand, by comparing the scales of acceleration involved in a general classical interaction and the characteristic time and length scales of elementary but realistic models of accelerated observers at a classical level, Mashhoon argued the necessity  to renounce to the {\it clock hypothesis}, the assumption that clocks measuring proper time parameters do not depend on the acceleration of the observer, in situations when radiation reaction effects are of relevance \cite{Mashhoon1990}. The clock hypothesis lies on the foundations of the description of relativistic spacetimes by means of Lorentzian structures \cite{Einstein1922, Pauli 1958}. Renouncing to the clock hypothesis in the above mentioned situations of strong radiation reaction regime opens the door for a non-trivial dependence of the proper time functional on the acceleration of the world line.

One class of spacetime metric structures with a dependence on acceleration can be motivated and constitute a fundamental ingredient of a proposed classical point charged particle dynamics. The development of a model of point charged particle in the framework of classical fields depending on the second jets of the probe particle world lines  \cite{Ricardo2012, Ricardo 2017, Ricardo2015} illustrated the possibility that classical fields are not independent physical objects defined over the spacetime manifold $M_4$. Instead, physical classical fields should be seen as geometric objects that depend on how they are probed by means of test particles. Once this hypothesis is adopted, it is natural to think that the spacetime metric depends upon higher order jet geometric objects too, expecting that a form of the equivalence principle will hold good in the new theory and also thinking that the metric will be related with the gravitational field in a similar way as in general relativity, fundamentally by relying on the principle of general covariance as heuristic principle.

In this context, the construction of consistent classical field-particle dynamical electrodynamic models was sketched and partially developed in \cite{Ricardo 2017}. The scheme of high order jet fields and metrics is a very amenable point of view that allows to accommodate radiation reaction in the case of classical electrodynamics in a consistent way with observation. Indeed, the classical fields considered by the theory depend upon how they are probed by test particles. Such amenability is realized by extra elements of the mathematical structure contained in the field that allow for the required purposes of having a dynamics for a point charged particle described by a second order equation without run-away and pre-acceleration solutions. Furthermore, the theory of spacetimes maximal proper acceleration implies, at least partially, the regularization of spacetime singularities of curvature type \cite{Ricardo 2020} and leads to a potential resolution of the black-hole information loss problem, since the existence of a maximal proper acceleration for the gravitational interactions implies that black-holes cannot totally evaporate and a finite remnant object must survive the process of evaporation \cite{Ricardo 2022c}.

High order jet fields and metrics are introduced by a procedure of deformation of lower order jet field structures. First, fields and metrics are defined on the spacetime manifold $M_4$ as it is usual in models formulated in the framework of Lorentzian spacetimes. Second, in the first step of the generalization, fields and metric structures are deformed to be defined on the first jet bundle manifold $J^1M_4$, very close to the situation that one finds for Finslerian spacetime models and Finslerian field generalizations \cite{Beem1, Brandt 1989, Vacaru Stavrinos Gaburov Gonta, Pfeifer Wohlfarth, GallegoPiccioneVitorio:2012, Javaloyes Sanchez 2020}. Despite the intrinsic differences between the first and second classes of models, where fields and structures live on $M_4$ or in $J^1M_4$ (in case of Finsler type structures, on sub-bundle of $TM_4$) respectively, differences that permeates the methodology of the corresponding differential geometric methods, these two classes of models are consistent with the clock hypothesis. The next step in the generalization process is to consider classes of spacetime geometric models that necessarily deviate from the principles of relativity theory, in particular, with respect to the clock hypothesis. One type of this class of models is the spacetimes of metric of maximal proper acceleration, which are almost conformably equivalent to Lorentzian spacetime models, but where the conformal factor depends upon the proper acceleration. These models can be seen as the simplest clock hypothesis violating general covariant models. The metric structures of maximal proper acceleration contain two absolute kinematic parameters, the speed of light in vacuum that fixes the causal structure, and a maximal proper acceleration $A_{\textrm{max}}$. On the other hand,  fields depend upon the world line position, velocity and acceleration of the test particles. A framework where the above characteristics are embedded is discussed in the following paragraphs.

\subsection{Spacetime of maximal proper acceleration as high order jet geometry}
Let us recall the fundamental concepts of the theory of metrics with maximal proper acceleration \cite{Ricardo2012, Ricardo2015, Ricardo 2024}. Our theory is a classical theory. Hence we assume that events are well described by points of a four-dimensional smooth manifolds $M_4$. If $\vartheta:I\to M_4$ is a smooth curve, then $\,^2 \vartheta (t)=\,\left(t,\vartheta^\mu(t),\frac{d{\vartheta}^\mu(t)}{dt},\frac{d^2{\vartheta}^\mu(t)}{dt^2}\right)$ determines a point of the second jet manifold $J^2M_4$. Note that we denote by $t$ one of the coordinates of $\,^2 \vartheta (t)$ and also the parameter of the curve $\vartheta:I\to M_4$ and similarly, when the curves are parameterized by other parameters.

Let us assume the existence of {\it co-moving ideal clocks}, regular devices that move with a probe particle and that are such that do not perturb significantly the evolution of the probe particle. The time laps measured with such clocks determine a functional $s$ from the aggregate of piecewise continuous curves $\vartheta:I\to M_4$ defined over the second jet manifold $J^2M_4$ to the reals $\mathbb{R}$. Such a functional is the {\it proper time functional}. Thus to the second jet lift $^2\vartheta:I\to J^2M_4$ of the curve $\vartheta:I\to M_4$, the proper time functional associates the number $s[\vartheta]\in\,\mathbb{R}$. By its attached significance,  $s[\vartheta]$ is an {\it scalar quantity} in the geometric sense: does not depend upon the coordinate system. Based on such property of invariance, the following construction appears natural. Let us consider an ideal clock $\mathcal{C}$ co-moving with the particle with world line $\vartheta:I\to M_4$. This means that the clock shares the same position, velocity and acceleration as the particle along the curve $\vartheta:I\to M_4$. The time measured by such a clock between two infinitesimally separated points of the curve is of the form $\Phi(\,^2\vartheta)(t)\,dt$. We further assume that the proper time functional is additive. In particular, the proper time functional can be expressed in the form
\begin{align*}
s[\vartheta]=\,\int_I\, \Phi(\,^2\vartheta)(t)\,dt
\end{align*}
where, in order to keep bounded expressions for the integral, $I$ is assumed to be a finite interval of $\mathbb{R}$.
Since $\Phi(\,^2\vartheta)(t)\,dt$ is an invariant, it can be identified with the value of a tensor acting on tangent vectors along $\vartheta(t)$. The simplest case is to consider that $\Phi$ is determined by a $2$-bilinear form acting on the tangent vector $\vartheta'$,
\begin{align*}
\Phi(\,^2\vartheta)(t)\,dt =\,\left( -g|_{\,^2\vartheta(t)}({\vartheta}',{\vartheta}')\right)^{1/2}\,dt ,
\end{align*}
where prime means derivative with respect to the parameter $t$ and where $g|_{\,^2\vartheta(t)}:T_{\vartheta(t)}M_4\times T_{\vartheta(t)}M_4\to \mathbb{R}$ is a bilinear, symmetric $2$-form living on the second jet $^2\vartheta (t)$. We add the further requirement for $g|_{\,^2\vartheta(t)}$ of being non-degenerate, which implies that it can be determined uniquely by the data $\{g|_{\,^2\vartheta(t)}(V,V),\quad V\in T_{\vartheta(t)} M_4\}$.
Hence the proper time functional can be expressed in the form
\begin{align}
s[\vartheta]=\,\int_I\,\left( -g|_{\,^2\vartheta(t)}({\vartheta}',{\vartheta}')\right)^{1/2}\,dt  .
\label{proper time}
\end{align}
Because the appearance of the second time derivatives in the base point $\,^2 \vartheta(t)\in\,J^2M_4$ where $g$ is evaluated, the proper time functional \eqref{proper time} is not re-parametrization invariant. Therefore, an appropriate choice of the parameter $t$ should be specified. To fix the construction, we proceed as follows.
We assume the dynamical situation where there is the possibility to define a coordinate system such that the second derivatives vanish along the curve $\vartheta :I\to M_4$,
\begin{align}
\frac{d^2\vartheta^\mu (\tilde{\tau})}{d\tilde{\tau}^2} =0,\quad \mu=0,1,2,3.
\label{free fall form of a geodesic}
\end{align}
The condition \eqref{free fall form of a geodesic} is not always possible to be satisfied covariantly in the form of a geodesic equation condition. It depends on the specific dynamical law followed by the test particle and on the choice of the parameter $\tilde{\tau}$. For instance, for a particle moving under the effect of an external magnetic field, only if the field is of forceless type, the Lorentz force equation can be re-cast covariantly as a geodesic equation admitting coordinate systems such that \eqref{free fall form of a geodesic} holds good \cite{BaoChernShen2000}. In such a case, the Berwald connection as defined in Finsler geometry is affine and the space is of {\it Berwald type} or, in the context of Lorentzian geometry, it is a generalized Berwald type as described in \cite{Ricardo 2017b}, for instance. When a choice of coordinates and time parameter is such that \eqref{free fall form of a geodesic} holds good, we can say that the test particle is in {\it free fall} and the coordinate system where the conditions \eqref{free fall form of a geodesic} hold good is a {\it free fall coordinate system}.

For a free fall particle with world line $\vartheta:I\to M_4$ in a free fall coordinate system, the {\it infinitesimal proper time} takes the form
\begin{align*}
ds =\,\psi(\tilde{\tau})\,d\tilde{\tau},\quad\quad \psi(\tilde{\tau})=\,\left(-g|_{(\vartheta(\tilde{\tau}), \dot{\vartheta} (\tilde{\tau}),0)}(\dot{\vartheta},{\dot\vartheta})\right)^{1/2} .
\end{align*}
 Under the constraints of being a free fall particle and if we described it by means of a free fall coordinate system, assuming homogeneity of degree zero in $g$ with respect to the velocities $\dot{\vartheta}^\mu$, the $1$-form $d\tau =\,\psi(\tilde{\tau})\,d\tilde{\tau}$ is invariant under re-parameterizations of $\tilde{\tau}$. This is because being the second derivatives equal to zero, there is no dependence in the base point in the second derivative variable. Hence the proper time \eqref{proper time} can be re-casted for these particular physical situations of free-fall as
\begin{align*}
s[\vartheta]=\,\int_I\,d{\tau}=\,\tau[\,^2\vartheta],
\end{align*}
which is re-parametrization invariant in the above sense.

By invoking the principle of general covariance, we argue that $d\tau$ must be a coordinate invariant object. That is, the type of situations when the relations \eqref{free fall form of a geodesic} hold with respect to a particular type of parameter ${\tau}$ determines a {\it physical situation}. In order to implement this point of view we need an adequate connection theory. Indeed, the equations \eqref{free fall form of a geodesic} partially determine the geodesic equations in free falling coordinates of a torsion free, Berwald type connection on $TJ^2 M_4$; the other conditions determining the Berwald connection are of {\it torsion type}. As it was argued in the context of Finslerian type geometries \cite{Ricardo 2017b}, a general class of structures consistent with the weak equivalence principle are generalized Berwald spacetimes. Such spacetimes are characterized by the property that their Berwald connection is defined on $M_4$, despite that  the spacetime metric structure lives on $TM_4$ or in an appropriate sub-bundle of maximal rank. Furthermore, it is need to interpret physically the type of physical situations that corresponds to such a mathematical description. Although no formally deduction, the construction resembles clearly the situation of the evolution of test particles in a pure gravitational field.

By identifying our notion of free falling system discussed above with the relativistic notion of free falling system and invoking a form of the Einstein equivalence principle by assuming that infinitesimally the gravitational effects do not have influence in the motion of point particles or any other form of non-gravitational phenomena (see \cite{Pauli 1958}, paragraph 51), if the $1$-form $d\tau$ measures the proper time, then it must be constructed from the special relativistic proper time invariant. Therefore, $d\tau$ is identified with the infinitesimal proper time of a relativistic Lorentzian spacetime metric $\eta$ and the relations \eqref{free fall form of a geodesic} with the auto-parallel curves of the associated Levi-Civita connection in free falling coordinates.

Let us consider the Lorentzian spacetime structure $(M_4,\eta)$ with signature $(-1,1,1,1)$ constructed in the above form. It describes the spacetime geometry when the test particles used to probe it are subjected only to the gravitational field.
The proper parameter of the Lorentzian metric $\eta$ is defined by the expression
\begin{align}
\tau [\vartheta] :=\,\int_{\vartheta} \,dt\,\left( -\eta({\vartheta}',{\vartheta}')\right)^{1/2} ,
\label{propertime eta}
\end{align}
where here prime notation means derivative with respect to the arbitrary parameter $t$.
The functional $\tau[\vartheta]$ is re-parametrization invariant. Note that to construct the derivative $\dot{\vartheta}$, it is enough that the world line $\vartheta:I\to M_4$ exists as a submanifold of $M_4$ and that $\eta$ is well-defined on $M_4$. At this point, we can already define the proper acceleration by the expression
$ a^2:=\,\eta(\nabla_{\dot{\vartheta}}\dot{\vartheta},\nabla_{\dot{\vartheta}}\dot{\vartheta}),$
 where $\nabla$ is the covariant derivative operator associated with the Levi-Civita connection of $\eta$.

Let us consider the simplest possible general covariant deformation of $\eta$ to a tensor defined over a second jet bundle. Such deformation is expressed in terms of an additional term that leads to a metric structure of the form
\begin{align}
g|_{\,^2\vartheta(\tau)} :=\,\left(1-\, \frac{\eta|_{\vartheta(\tau)}(\nabla_{\dot{\vartheta}(\tau)}\dot{\vartheta}(\tau),
\nabla_{\dot{\vartheta}(\tau)}\dot{\vartheta}(\tau)
)}{A^2 _{\textrm{max}}}\right)\eta|_{\vartheta(\tau)}
\label{maximalaccelerationmetric}
\end{align}
Although it is not re-parametrization invariant, this bilinear form acting on elements of the product $TM_4\times TM_4$, determines our model of high order jet geometry. We call it the {\it metric of maximal acceleration}. Although it is defined over the point $^2\vartheta(\tau)=\,\left(\tau,\vartheta^\mu(\tau),\frac{d{\vartheta}^\mu(\tau)}{d\tau},\frac{d^2{\vartheta}^\mu(\tau)}{d\tau^2}\right)\in\, J^2M_4$, its action on vector fields over $M_4$ is bilinear. $g$ is degenerated; the domain of degeneration of $g$ is
\begin{align*}
\mathcal{S}:=\left\{\,^2\vartheta(\tau)\in \,J^2M_4\,s.t.\quad  1-\, \frac{\eta|_{\vartheta(\tau)}(\nabla_{\dot{\vartheta}(\tau)}\dot{\vartheta}(\tau),
\nabla_{\dot{\vartheta}(\tau)}\dot{\vartheta}(\tau)
)}{A^2 _{\textrm{max}}}=0\right\} .
\end{align*}

In order to illustrate the dependence of the metric with the acceleration, let us consider the spacetime metric structure in the surrounding of a black hole. According to general relativity, the spacetime metric is determined by the mass, the charge and the angular momentum of the black hole. In contrast, in a spacetime of maximal proper acceleration, although $\eta$ is the Kerr-Newmann solution, the metric $g$  depends upon the type of particle used to probe it. The metric $g$ differs by a conformal factor depending on the acceleration when it is probed by means of charged particle like electrons or protons, rather than  neutrons.

The metric of maximal accelerations corresponds to a particular class of geometric object encoded in the following notion.
Let $J^k M_4$ be the $k$-jet bundle over $M$ and $\eta$ a pseudo-Riemannian metric on $M$.
\begin{definicion}
Let $pr:T^{(p,q)}J^k M_4\to J^kM_4$ be a tensor bundle over  $J^k M_4$.
A $k$-tensor field of type $(p,q)$  is a map $\hat{S}:I\to T^{(p,q)}J^kM_4$ such that for each lift $\vartheta :I\to M_4$ to $^k\vartheta: I\to J^k M_4$, the following diagram commutes:
\begin{align*}
\xymatrix{ &
{T^{(p,q)} J^kM_4} \ar[d]^{pr}\\
{ I} \ar[ur]^{\hat{S}}  \ar[r]^{^k\vartheta } & { J^kM_4}.}
\end{align*}

A generalized $k$-differential form along the curve $^k\vartheta :I\to M_4$ are similarly defined.
\label{generalized form and tensor}
\end{definicion}
Horizontal differential forms and covariant tensor fields are canonically defined from the projection $\pi_k:J^k M_4 \to M_4$. Recall that a vertical vector is an element of the kernel of $d\pi:TM_4\to M_4$. Vertical vector fields define a distribution of $TM_4$. An horizontal differential form is zero on the vertical distribution of $TM_4$ \cite{KolarMichorSlovak}. The metric of maximal proper acceleration is an horizontal covariant tensor of type $(0,2)$.

Given the timelike curve $\vartheta:I\to M_4$, $g$ determines the physical proper
time along $\vartheta$. It is given by
\begin{align}
s [\vartheta] :=\,\int_{\vartheta} \,d\tau\,\left( -g|_{\,^2\vartheta(\tau)}(\dot{\vartheta},\dot{\vartheta})\right)^{1/2} = \,\int_{\vartheta} \,d\tau\,\left(1-\frac{\eta(\nabla_{\dot{\vartheta}}\dot{\vartheta}(\tau),
\nabla_{\dot{\vartheta}}\dot{\vartheta}(\tau))}{A^2_{\textrm{max}}}\right)^{1/2},
\label{propertime g}
\end{align}
where the parametrization of the curve is taken with respect to the proper time $\tau$ of $\eta$.  $s [\vartheta]$ is the time measured by an ideal clock $\mathcal{C}:I\to M_4$ co-moving with a particle with world line $\vartheta:I \to M_4$ in the sense that $\,^2\mathcal{C}(\tau)=\,^2\vartheta(\tau)$.
Observe that for timelike curves,  $\tau [\vartheta]\geq s[\vartheta ]$.

\begin{comentario}
The Lorentzian metric $\eta$ is the limit of $g$ in the sense that $\eta=\,\lim_{A_{\textrm{max}}\to +\infty} g$. This limit does not depend upon the second jet $^2\vartheta$ where $g$ is evaluated, but only on the point $\vartheta(t)\in\,M_4$. Although physically different from the initial interpretation of $\eta$, this point of view is very useful for certain formal arguments.
\end{comentario}

The metric of maximal proper acceleration provides a generalization of the notion of angle by means of analogous expressions to the standard formulae for timelike vectors in Lorentzian spacetimes \cite{Oneill 1983}. Given two tangent vectors $V,W\in T_p M_4$, the  {\it hyperbolic angle} $\theta$ between the two vectors $V,W$ is determined by the expression
\begin{align}
\cosh_\eta (V,W) : =\,-\frac{\eta(V,W)}{|\eta(V,V)|^{1/2}\,|\eta(W,W)|^{1/2}}.
\end{align}
In a similar way, in a spacetime of maximal proper acceleration $(g,M_4)$, the hyperbolic angle between two four-vectors $V, W\in T_{\vartheta (0)}M_4$ evaluated at the second jet $^2\vartheta (0)\in J^2M_4$ determined by the curve $\vartheta:\tilde{I}\to M_4$ is
\begin{align}
\cosh_g (V,W) :=\, -\frac{ g_{^2\vartheta (0)}(V,W)}{|g_{^2\vartheta (0)}(\eta(V,V)|^{1/2}\,\,|g_{^2\vartheta (0)}(W,W)|^{1/2}}.
\end{align}
Then form the definition \eqref{maximalaccelerationmetric} of the metric $g$ it follows
\begin{align*}
\cosh_g (V,W) =\,-\frac{\left(1-\frac{a^2}{A^2_{\textrm{max}}}\right)\eta(V,W)}{\left(1-\frac{a^2}{A^2_{\textrm{max}}}\right)^{1/2}\,|\eta(V,V)|^{1/2}\,
\left(1-\frac{a^2}{A^2_{\textrm{max}}}\right)^{1/2}\,|\eta(W,W)|^{1/2}}= \cosh_\eta (V,W) ,
\end{align*}
where $a^2=\eta (\nabla_{\dot{\vartheta}}\dot{\vartheta},\nabla_{\dot{\vartheta}}\dot{\vartheta})$.

The {\it Newtonian angle} is defined by means of projection operators determined by world line of observers as follows \cite{BennTucker}. Given a Lorentzian metric $\eta$ and a timelike curve $\vartheta :I\to M_4$, there is a canonical projector orthogonal to $\dot{\vartheta}$,
\begin{align*}
\Pi_\eta (\vartheta) :T_{\vartheta} M_4 \to T_{\vartheta} M_4,\,\, Z\mapsto \Pi_\eta (\vartheta) Z=\, Z-\frac{\eta(\dot{\vartheta}, Z)}{\eta(\dot{\vartheta},\dot{\vartheta})}\,\dot{\vartheta},
\end{align*}
Then for two tangent vectors ${V},{W}\in T_{\dot{\vartheta(0)}}M_4$ at the point $p=\vartheta (0)$, the Newtonian angle is defined by the relation
\begin{align}
\cos_{\eta,\vartheta} (V,W) : =\,-\frac{\eta(\Pi_\eta(\vartheta ){V},\Pi_\eta(\vartheta ){W})}{|\eta(\Pi_\eta(\vartheta ){V},\Pi_\eta(\vartheta ){V})|^{1/2}\,\,|\eta(\Pi_\eta(\vartheta ){W},\Pi_\eta(\vartheta ){W})|^{1/2}}.
\label{Newtonian angle}
\end{align}
Let us consider the analogous projector operator $\Pi_g$.
It is not difficult to show that the projection operators defined by means of $g$ and $\eta$ coincide, $\Pi_g =\Pi_\eta$. The analogous definition of Newtonian angle using metric $g$ is
\begin{align}
\cos_{g,\vartheta} (V,W) : =\,-\frac{g(\Pi_g(\vartheta ){V},\Pi_g(\vartheta ){W})}{|g(\Pi_g{V},\Pi_g(\vartheta ){V})|^{1/2}\,|g(\Pi_g(\vartheta ){W},\Pi_g(\vartheta ){W})|^{1/2}}.
\label{Newtonian angle g}
\end{align}
Following an argument analogous to the case of hyperbolic angles just given, it follows the equality $\cos_{g,\vartheta} (V,W)=\,\cos_{\eta,\vartheta} (V,W)$.

These two equivalences between the notions of angles (hyperbolic and Newtonian) associated to $g$ and $\eta$ hold because $g$ and $\eta$ are conformably related by a non-zero conformal factor. However,  the equivalence between $g$ and $\eta$ is not true when $a^2 =\,A^2_{\textrm{max}}$, since in this case the conformal factor is zero.

\subsection{Upper bound in the maximal proper acceleration}
Although if the pace of ideal clocks could depend on the acceleration of the curve, it is a physical requirement that the causal character of a curve does not depend on the choice of $g$ and $\eta$. Otherwise, the concept of causal cone could depend upon the acceleration of curves. In order to implement this concept, let us consider a smooth curve $\vartheta:I\to M_4$ and assume that $\eta(\dot{\vartheta},\dot{\vartheta)}\neq 0$ and that it is parameterized by the proper time associated with $\eta$, $\eta(\dot{\vartheta},\dot{\vartheta})=\,-1$. Then the curve $\vartheta:I\to M_4$ parameterized by $\tau$ has the same causal character with respect to $g$ than $\eta$ iff the following condition meets,
 \begin{align}
  0 \leq \eta(\nabla_{\dot{\vartheta}}\dot{\vartheta},\nabla_{\dot{\vartheta}}\dot{\vartheta})<\,A^2_{\textrm{max}}.
  \label{boundedconditionforacceleration}
 \end{align}
Therefore, the upper bound \eqref{boundedconditionforacceleration} implies that $A_{\textrm{max}}$ is the maximal acceleration. Note that
\begin{align*}
g(\nabla_{\dot{\vartheta}}\dot{\vartheta},\nabla_{\dot{\vartheta}}\dot{\vartheta})=\,
\left(1-\,\frac{\eta(\nabla_{\dot{\vartheta}}\dot{\vartheta},\nabla_{\dot{\vartheta}}\dot{\vartheta})}{A^2_{\textrm{max}}}\right)
\eta(\nabla_{\dot{\vartheta}}\dot{\vartheta},\nabla_{\dot{\vartheta}}\dot{\vartheta})<\eta(\nabla_{\dot{\vartheta}}\dot{\vartheta},\nabla_{\dot{\vartheta}}\dot{\vartheta}).
\end{align*}
Thus the concept of maximal proper acceleration with respect to $\eta$ is compatible with the concept of maximal acceleration with respect to $g$.

Besides the case $A^2_{\textrm{max}}>\,\eta(\nabla_{\dot{\vartheta}}\dot{\vartheta}(\tau),\nabla_{\dot{\vartheta}}\dot{\vartheta}(\tau))$, there is another possibility that precludes the full equivalence of the causal structures associated with $g$ and $\eta$, corresponding to the  {\it world lines of maximal proper acceleration} \cite{Ricardo2015, Castro-Perelman 2022}. These are the world lines $\vartheta:I\to M_4$  such that
\begin{align*}
1- \frac{\eta(\nabla_{\dot{\vartheta}}\dot{\vartheta}(\tau),\nabla_{\dot{\vartheta}}\dot{\vartheta}(\tau))}{A^2 _{\textrm{max}}}=0,\quad\eta(\dot{\vartheta},\dot{\vartheta})\neq 0,
\end{align*}
rending the curve $\vartheta:I\to M_4$ a null curve with respect to $g$, while with respect to $\eta$ is timelike.

The value of the maximal acceleration is the least upper bound, but it is not a maximum value for the proper acceleration, since it is not attainable by a massive, physical particle.
Note that the value of the maximal proper acceleration $A_{\textrm{max}}$ is not specified in the above construction. Indeed, $A_{\textrm{max}}$ depends upon the type of dynamics involved in the test particle acceleration and could also depend upon the specie of particle.
\subsection{Maximal acceleration in high order jet electrodynamics}
Let us consider the theory of high order jet electrodynamics  \cite{Ricardo 2017,Ricardo 2024}. In such a theory, the electromagnetic field depends on the second jet of the test particles being used to test it, while the spacetime is of a spacetime of maximal proper acceleration. In particular, the electromagnetic field and the $4$-current are generalized forms of the class indicated in {\it Definition} \ref{generalized form and tensor}. Starting with a formal Lorentz force formulated in terms higher order jet fields and in the context of higher order jet electrodynamics, a formal argument in terms of $\epsilon =a^2/A^2_{\textrm{max}}$ argument exploiting the formal structure of the fields and requiring compatibility with a minimal generalized form of Larmor's covariant radiation law leads to a second order differential equation (see reference \cite{Ricardo 2017} or Appendix C in reference \cite{Ricardo 2024}). In general covariant language, such an equation of motion for a point charged particle with inertial mass $m$ and charge $q$ in an arbitrary frame is given by the expression
\begin{align}
m_0\,\nabla_{\dot{\vartheta}}\dot{\vartheta} =\,q\,\widetilde{\iota_{\dot{\vartheta}}F}-\,\frac{2}{3}\,{q^2}\,
g(\nabla_{\dot{\vartheta}}\dot{\vartheta},\nabla_{\dot{\vartheta}}\dot{\vartheta})\,\dot{\vartheta},
\label{equationofmotion}
\end{align}
where $F$ is the $2$-form Faraday form, $\iota_{\dot{\vartheta}}$ is the interior derivative with respect to $\dot{\vartheta}$ and $\widetilde{\iota_{\dot{\vartheta}}F}$ is the dual $1$-form determined by the metric $\eta$ of the  $1$-form ${\iota_{\dot{\vartheta}}F}$ \cite{BennTucker}.
This equation is formally the same than one appearing in Bonnor's theory \cite{Bonnor}, but differs on the interpretation of the inertial mass and in the geometric nature of fields and metric: while Bonnor's theory is formulated by means of standard fields and currents in a Lorentzian spacetime, our theory is based upon the notion of higher order fields and metric of maximal proper acceleration\cite{Ricardo 2024}. These fundamental differences lead to inequivalent dependence of the mass in terms of the proper acceleration squared in Bonnor's theory and our theory. On the other hand, one can also consider the relation of our theory with the well known Landau-Lifshitz equation, also a second order differential equation \cite{Landau Lifshitz 2}. It is quite direct that both equations appear quite different. This is natural, since they have rather different origin and motivation: while Landau-Lifshitz equation is established from the third order Lorentz-Dirac equation by a reduction of order procedure, in our theory there is not involved any third order differential equation at any stage.

The equation \eqref{equationofmotion} leads to an upper bound for the particle acceleration \cite{Ricardo2015,Ricardo 2019,Ricardo 2024}. By contracting with $g$ each side of the \eqref{equationofmotion}, we obtain
 \begin{align*}
 m^2_0\left(1-\frac{a^2}{A_\textrm{max}}\right)\,a^2\,= \,F^2_L +\,\left(\frac{2}{3}\,q^2\right)^2\,(a^2)^2\,g(\dot{\vartheta},\dot{\vartheta}).
 \end{align*}
 where $F^2_L =\, q^2\, g(\widetilde{\iota_{\dot{\vartheta}}F},\widetilde{\iota_{\dot{\vartheta}}F})\geq 0$.
Assuming that at first approximation the acceleration is given by the Lorentz force, it follows that
\begin{align}
 A^2_{\textrm{max}}\leq \,\left(\frac{3}{2}\,\frac{m_0}{q^2}\right)^2 .
\label{valueofthemaximalacceleration}
\end{align}
Since this is the only acceleration scale in the theory, we assume that it is indeed the value of the maximal proper acceleration for high order jet electrodynamics.

The non-homogeneous dependence $m/q^2$ in the expression for the  maximal proper acceleration \eqref{valueofthemaximalacceleration} was exploited in \cite{Ricardo 2019,Ricardo 2024} to show that the effective maximal proper acceleration of a bunch of particles is drastically dropped to scales that can allow experimental signals. The model for the bunch of particles that we use consists of a point particle with the whole mass and charge of the bunch. If the bunch contains $N$ individual charged particles, then the total charge is $N\,q$ and the total inertial mass is $N\,m$. It was argued in \cite{Ricardo 2024} that within the conditions of stability of the bunch, the bunch can be described as a point charged particle with mass and charge $(N\,m, N\,q)$ whose world lines are solutions of the equation \eqref{equationofmotion}. The equation of motion \eqref{equationofmotion} can be treated perturbatively in terms of the quotient $\epsilon=a^2/A^2_{\textrm{max}}$, with the leading term being the Lorentz force term. Consequently, at zero order in $\epsilon$, the model for a charged particle is equivalent to the Lorentz force equation. Assuming this, the acceleration $|a|$ depends upon the quotient $m_0/q$, while the maximal proper acceleration  $A_{\textrm{max}}$ depends on the quotient $m_0/q ^2$. Thus under the influence of the external fields of the accelerator and for acceleration squared $a^2 $ small enough compared with $A^2_{\textrm{max}}$, the bunch of particles will have the same acceleration as an individual point charged particle with charge and mass $(m_0, q)$.

Besides disregarding all the effects of beam dynamics related with the composition, size and structure of the bunch, especially disregarding space charge effects, the model of the bunch as a sole particle breaks down when
\begin{align*}
\big| F^2_L \big|\sim \,\big|\left(\frac{2}{3}\,q^2\right)^2\,(a^2)^2\,g(\dot{\vartheta},\dot{\vartheta})\big |,
\end{align*}
a condition that can be re-casted in a closer form:
\begin{align}
\left(\frac{q}{m_0}\right)^2\,\big| g(\widetilde{\iota_{\dot{\vartheta}}F},\widetilde{\iota_{\dot{\vartheta}}F})\big|\sim \,\frac{a^2}{A^2_{\textrm{max}}}\,a^2,
\label{consistence condition of the model}
\end{align}
where we have used the condition $\eta(\dot{\vartheta},\dot{\vartheta})=\,-1$. The left side of this condition does not depend upon the number population $N$. Thus, as long as we are far from reaching the condition \eqref{consistence condition of the model} and other dynamical effects are disregarded, the approximation of treating the bunch as a single charged particle with mass $N\, m_0$ and charge $N\, q$ remains valid.

Let us assume that we are in the dynamical regime where we can use the model exposed above. The maximal proper acceleration of a stable bunch composed by $N$ particles is reduced with respect to \eqref{valueofthemaximalacceleration} by a factor $N$,
\begin{align}
 A^2_{\mathrm{max}}(N)=\,\frac{1}{N}\,\left(\frac{3}{2}\,\frac{m_0}{q^2}\right)^2 .
 \label{valueofthemaximalacceleration2}
\end{align}

\subsection{Other theories showing maximal proper acceleration}
In order to illustrate further the concept of maximal proper acceleration, let us consider two distinguished examples of frameworks leading to upper bounds in the acceleration. Further examples can be found in \cite{Ricardo Nicolini 2018}.
Let us consider the Schwinger pair creation of particle-antiparticle in a very strong electric field and the associated acceleration scale $A_{S}$. The creation of a pair of particles would imply an energy gain at least equal to $2\,m\,c^2$ in the initial rest frame, where the length scale of the system is associated with the Compton wavelength and by system we include also the pair particle-antiparticle generated. In such a process of energy  gain, the speed can change maximally from zero to nearly equal to $c$ in a time equal to $(h/mc) / c$. The corresponding acceleration scale is then bounded by $A_{S}\sim 4\pi\, m c^3 / h$. As a reference, for an electron, the scale of this acceleration is of order $A_S\sim 2\,\cdot 10^{29} m/s^2$, an order of magnitude lower than the maximal acceleration of higher order jet electrodynamics.

Similar considerations apply to Caianiello's quantum mechanical derivation of the maximal proper acceleration \cite{Caianiello 1984}. The starting point is an uncertainty relation of the form
\begin{align*}
\Delta E \Delta f(t)\geq \,\frac{\hbar}{2}\, \left| \frac{df}{dt}\right|.
\end{align*}
where here $t$ refers to a time attached to the observer measuring the energy $E$ and the function $f$.
If $\Delta E\leq \,mc^2$, $f=v$ and the relativistic constraint $\Delta v\leq c$ holds, then the proper acceleration $a$ must be bounded by a maximal value given by
\begin{align}
A_{C}=\,2\,\frac{mc^3}{\hbar}.
\label{Maximal acceleration of Caianiello 2}
\end{align}

Note that the arguments leading to Schwinger's and Caianiello's maximal acceleration are of heuristic nature and only demonstrate the existence of certain scales of acceleration. This is because, according to the orthodox  interpretation of quantum mechanics, at the quantum level description there is no notion of classical acceleration of the way discussed in this paper, ready to be used in jet geometry. Other interpretations of quantum mechanics. Bohm-de Broglie theory \cite{Bohm}, stochastic quantum mechanics \cite{Nelson} or emergent quantum mechanics \cite{Ricardo2014} may provide a framework to speak of acceleration at the quantum level, but we are not considering these issues in this paper.
\subsection{Covariant momentum in spacetimes of maximal proper acceleration}
Particles {\it on-shell} are partially characterized by the associated four momentum. Here we generalized the construction of the notion of four momentum as discussed in \cite{BennTucker} to the description of point charge particles in spacetimes of maximal proper acceleration. Let us consider a timelike curve $\vartheta:I\to M_4$ that will represent the world line of the particle. It will be parameterized by the proper time according to the limit metric $\eta =\, \lim_{A_{\textrm{max}}\to +\infty} g =\,\eta$. Let us consider a vector field $P:I\to M_4$ along $\vartheta$ such that
\begin{align}
g_{^2\vartheta}(P,P)=\,-m^2\,c^4.
\label{on-shell condition}
\end{align}
$P$ is the four-momentum of the particle with world line $\vartheta:I\to M_4$; $m$ is the inertial mass of the particle.
Let us consider now an observer, that is modeled by a timelike vector field ${\dot{\mathcal{O}}}\in\,\Gamma TM$. The vector field $\dot{\mathcal{O}}$ has associated integral curves of the form $\mathcal{O}:I'\to M_4$ and it also determines a vector field along $\vartheta:I\to M_4$ by pointwise identification. Such a vector field along $\vartheta$ is modified, if appropriate, to be unitary in the sense that $\eta({\dot{\mathcal{O}}}, {\dot{\mathcal{O}}})=\,-1$. With respect to ${\dot{\mathcal{O}}}$, there is an unique decomposition ${P}=\,c^2\,\mathcal{P}+\,\mathcal{E}\,{\dot{\mathcal{O}}},$
 where $g_{^2\vartheta}(\mathcal{P},{\dot{\mathcal{O}}})=\,\eta(\mathcal{P},{\dot{\mathcal{O}}})=0$. Then we have that
\begin{align*}
g_{\,^2\vartheta}(P,P)& =\,c^2\,g_{^2\vartheta}(\mathcal{P},\mathcal{P})+\,\mathcal{E}^2\,g_{\,2\vartheta}(\dot{\mathcal{O}},\dot{\mathcal{O}})=\\
& =\,\left(1-\frac{a^2}{A^2_{\textrm{max}}}\right)\,\Big[c^2\,\eta(\mathcal{P},\mathcal{P})-\,\mathcal{E}^2\Big]=\,-m^2\,c^4 ,
\end{align*}
where $a^2=\,\eta(\nabla_{\dot{\vartheta}}{\dot{\vartheta}}, \nabla_{\dot{\vartheta}}{\dot{\vartheta}})$ is the proper acceleration squared of the classical particle.
The {\it Newtonian velocity} is defined, as for the relativistic case, by the relation $\mathcal{V} :=\,\frac{\mathcal{P}}{\mathcal{E}}\,c^2$. Note that, according to these definitions, the Newtonian velocity and the three momentum $\mathcal{P}$ are spacelike four vectors.

The on-shell relation \eqref{on-shell condition} leads to the relation
\begin{align*}
\mathcal{E}^2 \,\frac{\eta(\mathcal{V},\mathcal{V})}{c^2}-\,\mathcal{E}^2 =\,-\,m^2\,c^4\,\left(1-\frac{a^2}{A^2_{\textrm{max}}}\right)^{-1} .
\end{align*}
Given an observer, the {\it energy} $\mathcal{E}$ and the Newtonian momentum $\mathcal{P}$ observed by  $\dot{\mathcal{O}}$ are given by the following expressions,
\begin{align}
\mathcal{E}(\tau)=\,\frac{1}{\sqrt{1-\frac{a^2(\tau)}{A^2_{max}}}}\,\frac{1}{\sqrt{1-\frac{\mathcal{V}^2(\tau)}{c^2}}}\,m\,c^2,
\label{modifiedE}
\end{align}
\begin{align}
\mathcal{P}(\tau)=\,\frac{1}{\sqrt{1-\frac{a^2(\tau)}{A^2_{max}}}}\,\frac{1}{\sqrt{1-\frac{\mathcal{V}^2(\tau)}{c^2}}}\,m\,\mathcal{V}(\tau),
\label{threemoment}
\end{align}
where here $a^2$ is the proper acceleration squared of the particle. Similarly as in a relativistic theory for massive point particles reaching the light cone domain, the work necessary to reach the maximal acceleration domain is divergent.

One particular type of co-moving observers is remarkably useful. An instantaneous co-moving observer $\dot{\mathcal{O}}$ at the instant $\tau$ with respect to a world line $\vartheta:I\to M_4$ is a co-moving observer in the sense that $^2\vartheta =\,^2\mathcal{O}$ such that the Newtonian velocity observed by $\dot{\mathcal{O}}$ is zero.
The rest mass of the particle is defined as the ratio $\mathcal{E}/c^2$ when the observer is instantaneously at rest with the particle, meaning $\mathcal{V}=0$. Thus the rest mass is given by the expression \cite{Ricardo2015,Ricardo 2024},
\begin{align}
\tilde{m} =\,\frac{m}{\sqrt{1-\frac{a^2}{A^2_{\textrm{max}}}}},
\label{modified rest mass}
\end{align}
where $a^2$ is the squared proper acceleration of the particle of world line $\vartheta: I\to M_4$.
The notions of rest mass and inertial mass are different in a theory of spacetimes with maximal proper acceleration than in a Lorentzian theory, a fact with potential relevant phenomenological consequences \cite{Ricardo 2019,Ricardo 2024}. This is consistent with the notion of inertial mass in Hamilton-Randers theory \cite{Ricardo2014,Ricardo2023}.

The on-shell condition \eqref{on-shell condition} dispersion relation for a classical particle is formally the standard relativistic dispersion relation,
\begin{align}
-\mathcal{E}^2 +c^2 \,\mathcal{P}^2 =\,-\, \tilde{m}^2 \, c^4 .
\label{dispersion relation}
\end{align}
where $\mathcal{E}$ and $\mathcal{P}$ are given by the expressions \eqref{modifiedE} and  \eqref{threemoment}, but where the rest mass $\tilde{m}$ is given by the expression \eqref{modified rest mass}. In the absence of acceleration, the relation \eqref{dispersion relation} coincides with the standard relativistic dispersion relation.

The mass at rest of a particle with inertial mass $m$ observed by the observer with world line  $\mathcal{O}:I'\to M_4$ is defined by the expression
\begin{align}
\tilde{m}|_{\,^2\mathcal{O}}:=\,\frac{m}{\sqrt{1-\frac{\eta(\nabla_{{\dot{\mathcal{O}}}}{\dot{\mathcal{O}}}
(\tau),\nabla_{{\dot{\mathcal{O}}}}{\dot{\mathcal{O}}}(\tau))}{A^2_{\textrm{max}}}}} .
\label{mass measured by O}
\end{align}
The mass at rest observed by $\dot{\mathcal{O}}$ differs from the mass at rest of the particle as given by the relation \eqref{modified rest mass}, since the proper acceleration that appears in $\tilde{m}|_{\,^2\mathcal{O}}$ is the proper acceleration squared of the observer $\dot{\mathcal{O}}$, instead of the proper acceleration squared of the particle. However, when the world line of the observer $\dot{\mathcal{O}}$ coincides with the world line of the particle, the two notions of mass given by the observer mass $\tilde{m}|_{\,^2\mathcal{O}}$ and the particle rest mass $\tilde{m}$, coincide. An advantage of the notion of observed mass $\tilde{m}|_{\,^2\mathcal{O}}$  is that it is applicable to non-classical particles or states, since there is no dependence on the location or jets associated with the particle of mass $m$. In particular, the notion is applicable to the excitations associated to quantum fields. We will use this notion in the next {\it section} on development of the Fulling-Davies-Unruh effect in spacetimes of maximal proper acceleration.
\section{On the Unruh temperature in spacetimes of maximal proper acceleration}\label{On the Unruh effect in spacetimes of maximal proper acceleration}

\subsection{Unruh temperature for hyperbolic observers in spacetimes of maximal proper acceleration}
An hyperbolic observer is a timelike vector field $\dot{\mathcal{O}}$ such that the squared proper acceleration $a^2(\tau)=\,\eta\left(\nabla_{{\dot{\mathcal{O}}}}\dot{\mathcal{O}}(\tau),\nabla_{{\dot{\mathcal{O}}}}{\dot{\mathcal{O}}}(\tau)\right)$ is constant along each of its integral curves and does not depend upon the particular integral curve. We discuss in this {\it section} the generalization of the Unruh temperature formula  \cite{Fulling,Davies,Unruh 1976} to hyperbolic observers in a spacetime of maximal proper acceleration.
The spacetime background that we consider is a flat spacetime in the sense that the limit of the metric structure $g$ when the parameter $A_{\textrm{max}}\to +\infty$ is a flat metric in the pseudo-Riemannian sense. Therefore, the metric of maximal proper acceleration that an hyperbolic observer experiences is the Minkowski metric multiplied by a constant factor,
\begin{align*}
g_{|^2\mathcal{O}}=\,\left( 1-\frac{a^2}{A^2_{\textrm{max}}}\right)\,\eta,
\end{align*}
 where $a^2:=\,\eta(\nabla_{\dot{\mathcal{O}}}\dot{\mathcal{O}}(\tau),\nabla_{\dot{\mathcal{O}}}\dot{\mathcal{O}}(\tau))$ is the acceleration of an observer with world line $\mathcal{O}:I\to M_4$.

Since the conformal factor relating $g_{|^2\mathcal{O}}$ with the Minkowski metric $\eta$ is constant, it is possible to transport directly the geometric statements that depend only on the Levi-Civita connection and Riemann curvature from Lorentzian geometry to the setting of spacetime metrics with maximal proper acceleration (see for instance \cite{Oneill 1983}, pg. 92 for the behaviour of different geometric quantities under nomothetic transformations). However, it is not clear how to generalize the notion of temperature as it appears in the standard Unruh temperature formula. Therefore, the possibility of generalizing the Unruh temperature formula to the new context of hyperbolic observers in $(M_4,g)$ should be analyzed in certain detail. Such generalization can be achieved by a direct method, that is, by developing a quantum field theory in spacetimes of maximal proper acceleration for different types of fields and developing a consistent notion of detector or observer in such a context. Alternatively, the indirect way to address the problem is to identify the theoretical elements required to establish the Fulling-Davies-Unruh effect in the standard theory. If such elements are present in the new theory and can be used in a similar way, then the Fulling-Davies-Unruh effect will still hold for the spacetimes under consideration. We choose to follow the second method.

Let us consider the action functional for a massive, scalar field formulated in a spacetime with metric of maximal acceleration $g=\,(1-a^2/A^2_{\textrm{max}})\,\eta$, where the proper acceleration squared $ a^2$ of the classical observer $\mathcal{O}:I\to M_4$ is constant. For hyperbolic observers, the constancy of the proper acceleration implies that if $(M,\eta)$ is a hyperbolic spacetime, then $(M,g)$ is hyperbolic too. Moreover, for hyperbolic observers, the metrics $\eta$ and $g$, differing just by the constant factor $(1-a^2/A^2_{\textrm{max}})$, have formally the same standard Klein-Gordon equations.
In order to show this las statement in detail, let us consider  the action of a scalar field.
In a spacetime of maximal acceleration we propose that the action for a neutral real scalar field is given by the expression
\begin{align}
\mathcal{A}[\phi,g]=\,\int d^4x \sqrt {- \det (g|_{\,^2\mathcal{O}})}\,\frac{1}{2}\, \left(g^{-1}|_{\,^2\mathcal{O}}(\nabla\phi, \nabla\phi)-\,\tilde{m}^2|_{\,^2\mathcal{O}}\,\phi^2\right) .
\label{action scalar field}
\end{align}
The domain of the action functional \eqref{action scalar field} must be restricted to the sub-maximal accelerations, $J^2 M_4\setminus \mathcal{S}$. This is an open set in the usual topology of $J^2 M_4$. Hence one can perform variational calculus in such a domain. Since the observer $\dot{\mathcal{O}}$ perceives the metric $g|_{^2\mathcal{O}}$, it is with respect to this metric that the action integral for fields must be defined. Also, let us remark that since the mass coefficient appearing in the action is the mass of the particle perceived by the observer $\dot{\mathcal{O}}$, the action
$\mathcal{A}[\phi,g]$ is numerically equal to the standard action of a scalar field in Minkowski space $\mathcal{A}\mathcal[\phi,\eta]$. Therefore, $\mathcal{A}[\phi,g]$ is a scalar that does not depend upon the observer $\dot{\mathcal{O}}$.

Being the proper acceleration constant, the critical condition $\delta \mathcal{A}= 0$ leads to the standard Klein-Gordon equation for particles of the mass $\tilde{m}^2|_{\,^2\mathcal{O}}$,
\begin{align}
\left(\Box_g -\tilde{m}^2|_{^2\mathcal{O}}\right) \phi=\,0.
\label{kleingordon equation for phi in g}
\end{align}
This equation is equivalent to the Klein-Gordon equation for a particle of mass $m$ in the spacetime $(M_4,\eta)$ regardless of the coordinate system used, because $\Box_g =\, (1-\frac{a^2}{A^2_{\textrm{max}}})^{-1}\,\Box_\eta$:
\begin{align*}
\left(\Box_g -\tilde{m}^2|_{^2\mathcal{O}}\right) \phi=\,\left(\Box_g -m^2 (1-\frac{a^2}{A^2_{\textrm{max}}})^{-1}\right) \phi =\,(1-\frac{a^2}{A^2_{\textrm{max}}})^{-1}\left(\Box_\eta -m^2 \right)\phi=0,
\end{align*}
where $\Box_g$ (resp. $\Box_\eta$) is the $4$-dimensional Laplacian of the metric $g$ (resp. the Laplacian of the metric $\eta$).

An inner product in the space of solutions of the Klein-Gordon equation \eqref{kleingordon equation for phi in g} is defined as follows. Let us consider a Cauchy hypersurface $\Sigma$ of $(M_4,\eta)$. For the case of flat spacetimes, it is an hypersurface whose normal unit vector is the timelike vector $\partial_t$. For each point of $\Sigma$ there is only one integral line associated to the observer field $\dot{\mathcal{O}}$ that passes through it. Then $g$ is evaluated in $\Sigma$  by evaluating the metric on each of the corresponding second jets $^2\mathcal{O}$ of the integral line of the observer. Then the scalar product is defined by
\begin{align}
(\phi_1,\phi_2)_g :=\,-\imath \int_\Sigma d^3 x \sqrt{\det g_3}\, g^{-1}(\tilde{n}, \phi_1 \nabla\phi^*_2-\phi^*_2\nabla\phi_1),
\label{scalar product for g}
\end{align}
where $\Sigma$ is a Cauchy hypersurface for both, for $g$ and for $\eta$, and $g_3$ and $n$ is the normal vector to $\Sigma$, $n$ being unitary in the sense that $g(n,n)=1$ and $\tilde{n}$ is the dual $1$-form associated to $n$. It is understood that the metric quantities are evaluated at jets $^2\mathcal{O}$ over points of $\Sigma$.

The scalar product \eqref{scalar product for g} is isometric to the Hilbert product determined by $\eta$,
\begin{align*}
(\phi_1,\phi_2)_g & =\,-\imath \int_\Sigma d^3 x \sqrt{\det g_3}\, g^{-1}(\tilde{n}, \phi_1 \nabla\phi^*_2-\phi^*_2\nabla\phi_1)\\
& =\,\,-\imath \int_\Sigma d^3 x \sqrt{\det\eta_3}\, \eta^{-1}(\widetilde{n'}, \phi_1 \nabla\phi^*_2-\phi^*_2\nabla\phi_1)=\,(\phi_1,\phi_2)_\eta ,
\end{align*}
where $g_3$, $\eta_3$ are the metrics on $\Sigma$ induced by $g$  and $\eta$ respectively, $n'$ is the normal unitary vector to $\Sigma$ such that $\eta(n',n')=1$, while $\widetilde{n'}=\,\eta(n',\cdot)$ is its dual $1$-form determined by $\eta$.

In the case of hyperbolic observers  revise if the requirements to derive the standard Unruh temperature relation hold good \cite{Alsing Milonni,Lambert}. For the case of hyperbolic observers in a spacetime of maximal acceleration, the following points are direct generalizations from the sufficient conditions to derive the Unruh temperature formula in flat spacetime:
\begin{enumerate}

\item \label{Claim 1} The dispersion relation is the relativistic dispersion relation.

\item \label{Claim 2} For hyperbolic spaces, the metrics $g$ and $\eta$ share a common timelike Killing vector, namely, the vector field $K=\partial_t$.

\item \label{Claim 3} For hyperbolic observers, since the proper acceleration is constant, the canonical commutation relations for the creation and annihilation operators hold good for fixed of the proper time parameter $s$ of $g$ and $\tau$ of $\eta$.

\item \label{Claim 4} For hyperbolic observers the Klein-Gordon equation in a given coordinate system are formally the same than for standard Lorentzian case for such coordinate systems.

\item \label{Claim 5} The solutions of scalar field equations in the form of positive frequency modes and negative frequency modes are formally the same than for the standard fields defined in Minkowski spacetime.

\end{enumerate}

Let us discuss the validity of the of claims above. Claim $(\ref{Claim 1})$ follows by considering plane wave solutions to the Klein-Gordon type equation \eqref{kleingordon equation for phi in g}. Effectively, if $\phi =\,\phi_0\,\exp (k_\mu x^\mu)$, then we have the dispersion relation
\begin{align}
-k^2_0+\vec{k}^2 =\,m^2.
\label{relativistic dispersion relation}
\end{align}
This is the standard relativistic dispersion relation for a particle of mass $m$. Let us remark that there is no contradiction between this relativistic form of the dispersion relations and the on-shell condition for a classical particle \eqref{on-shell condition}, because the dispersion relation \eqref{relativistic dispersion relation} refers to systems in the asymptotic limit where plane wave applies. This limit is compatible with free system, that classically translates into the condition of zero proper acceleration $a=0$. Thus assuming the standard relation between $P=\,\hbar\,k$, there is compatibility between the relativistic dispersion relation and the on-shell condition \eqref{on-shell condition} in this limit. On the other hand, the relation \eqref{on-shell condition} applies to particles with proper acceleration and it also has a classical derivation and interpretation. Note that in the limit of validity of the dispersion relation \eqref{relativistic dispersion relation}, the observer $\mathcal{O}$ can have acceleration.

Claim $(\ref{Claim 2})$ follows directly because the proper acceleration $a$ of the observer $\mathcal{O}$ is constant and therefore, there is a constant scale factor between $g$ and $\eta$.

Claim $(\ref{Claim 3})$ follows from the homogenous, constant scale factor between $s$ and $\tau$ when the observer is hyperbolic. Then they determine the same $t$-time slits hypersurfaces. Equal time commutation relations are then the canonical ones.

Claim $(\ref{Claim 4})$ has been proved above and claim $(\ref{Claim 5})$ follows from claim $(\ref{Claim 4})$.

Therefore, for hyperbolic observers, the changes in the assumptions in our theory with respect to the standard theory of the Fulling-Davis-Unruh effect in flat spacetime are formally minimal.

We are now in position to generalize the Unruh temperature formula to hyperbolic observers in spacetimes of maximal proper acceleration.
Due to the form of the dispersion relation \eqref{relativistic dispersion relation}, the Fourier analysis over $\mathbb{R}^4$ (or over $\mathbb{R}^2$, if we are consider the 1+1 case) is the same than for a Lorentzian spacetime. In the case of an hyperbolic observer with world line $\mathcal{O}:I\to \mathbb{R}$, the field equations for scalar fields in a spacetime of maximal proper acceleration coincide with the standard Klein-Gordon equation. Furthermore, the Hilbert spaces of scalar fields associated with the spacetimes $(\mathbb{R}^4,g)$ and $(\mathbb{R}^4,\eta)$ are isometric and the notion of hyperbolic observer is the same for $\eta$ and for $g$. Then it is clear that an hyperbolic observer in a spacetime of maximal proper acceleration will detect a thermal bath at the same temperature than an hyperbolic observer with the same proper acceleration in the Minkowski spacetime and that the Unruh temperature observed by $\mathcal{O}$ is given by the standard Unruh temperature formula \cite{Davies,Unruh 1976},
  \begin{align}
 T=\, \frac{\hbar\, a}{2\,\pi\,c\,k_B},
 \label{temperature}
 \end{align}
where $a^2=\,\eta(\nabla_{\dot{\mathcal{O}}}\dot{\mathcal{O}},\nabla_{\dot{\mathcal{O}}}\dot{\mathcal{O}})$ is the proper acceleration of the observer world line $\mathcal{O}:I\to M_4$.

The maximal acceleration limit is a formal one. Therefore, the relation \eqref{temperature} is well defined in the interval $[0, A_{\textrm{max}})$.

Once the Unruh temperature formula has been extended to hyperbolic observers in a spacetime of maximal acceleration in the domain $J^2M_4\setminus S$, the existence of a maximal Unruh temperature as a formal upper bound emerges directly: the existence of a maximal proper acceleration $A_{\textrm{max}}$ implies the existence of a maximal Unruh temperature,
 \begin{align}
 T_{\textrm{max}}=\, \frac{\hbar\, A_{\textrm{max}}}{2\,\pi\,c\,k_B},
 \label{maximal temperature}
 \end{align}
 that must be considered as a formal limit. The analogous relation \eqref{maximal temperature} was assumed in other contexts by Brandt \cite{Brandt1983} and by Caianiello and Landi \cite{CaianielloLandi} in different contexts than here.

One direct consequence is that the mass at rest $\tilde{m}|_{^2\mathcal{O}}$ observed by $\mathcal{O}$ can be seen as a function of the temperature of the thermal bath observed by $\mathcal{O}$, due to the relation \eqref{temperature},
\begin{align}
\tilde{m}|_{^2\mathcal{O}}=\,\frac{m_0}{\sqrt{1-\frac{T^2}{T^2_{\textrm{max}}}}}.
\label{mass in term of temperature}
\end{align}
This result resembles, but is essentially different, from the result by Castro-Perelman  in the context of Born reciprocity relativity theory \cite{Castro-Perelman 2024}.
\section{The Unruh temperature in laser-plasma acceleration for high order jet classical electrodynamics}\label{The Unruh temperature in laser-plasma acceleration for high order jet classical electrodynamics}
The Fulling-Davies-Unruh effect establishes an equivalence between temperature and acceleration of the form
\begin{align*}
T \approx \, 4.05\cdot 10^{-21}\,K \,m^{-1}\,s^2\cdot \,a[ms^{-2}],
\end{align*}
where the acceleration is measure in $ms^{-2}$.
 We attach a classical accelerated observer to the particle being accelerated. If according to our model of classical point charged particle, the bunch of particles being accelerated is considered as a sole compact, stable dynamical system containing $N$ identical charged particles of mass and charge $(m, q)$, then the bunch will experience accelerations that are bounded by the expression \eqref{valueofthemaximalacceleration2}. As a consequence of the relation \eqref{maximal temperature}, the attached observer perceives a thermal bath whose temperature is bounded by a {\it maximal Unruh temperature} of the form
 \begin{align}
 T_{\textrm{max}}(N)=\, \frac{1}{N}\frac{\hbar}{2\,\pi\,c\,k_B} \left(\frac{3}{2}\,\frac{m}{q^2}\right)^2  .
 \label{maximal temperature in electrodynamics}
 \end{align}

The model of electrodynamics discussed above has striking consequences for the phenomenology of laser-plasma acceleration, since for these systems, the maximal acceleration of the bunch decreases with the factor $1/N$ \cite{Ricardo 2019, Ricardo 2024}. For a bunch of particles containing $N=\,10^{8}$ being accelerated by the wake-field mechanisms, the damping in the scale by the factor $1/N$ implies a maximal proper acceleration  for the bunch of order $10^{22}\,m/s^2$. Hence the maximal temperature of the thermal bath that an hyperbolic observer can feel in wake-field acceleration with $N=\,10^{8}$ attached to the bunch is of order $T_{\textrm{max}}\sim\,10^2\,K$, in contrast with the unbounded Unruh temperature predicted by the standard Fulling-Davies-Unruh effect associated with Maxwell-Lorentz electrodynamics and other models of electrodynamics with unbound proper acceleration. We can also compare \eqref{maximal temperature in electrodynamics} with he maximal Unruh temperature for a single elementary charged particle in high order jet electrodynamics,
\begin{align}
T_{\textrm{max}}(N=1)=\, \frac{\hbar}{2\,\pi\,c\,k_B} \left(\frac{3}{2}\,\frac{m}{q^2}\right)^2 ,
 \label{maximal temperature in electrodynamics N=1}
 \end{align}
 For an electron, this temperature is of order $T_{\textrm{max}}(e)\sim \,10^{10} \,K$, while for a proton is of order $T_{\textrm{max}}(p)\sim \,10^{13} \,K$.
 Increasing the bunch population $N$ leads to a reduction of the maximal Unruh temperature according to the inverse of the population. However, space charge effects that affect the stability of very large bunches precludes the limit $N\to +\infty$.

\section{Conclusion and Discussion}\label{Discussion}
 The introduction of the concept has been motivated in a significatively different way with respect to previous presentations \cite{Ricardo2007,Ricardo2015,Ricardo 2024}. The emphasis of the new present presentation is on the constructibility of the proper time and the underlying Lorentzian metric $\eta$ by implementing general covariance and a generalized form of Einstein's equivalence principle in the setting of high order jet geometry. We have seen that $\eta$  corresponds to the case of probing particles not being accelerated. This corresponds to the interaction of the point particle with the pure gravitational field. Our treatment address the question of how a charged particle evolves in a gravitational field, that should be according the geodesic equation of $\eta$, which coincide with the geodesics of the Berwald connection for $g$ with the same non-trivial connection coefficients than the Levi-Civita connection of $\eta$. Second, we have discussed the extension of the Unruh temperature formula for hyperbolic observers in spacetimes of maximal proper acceleration, leading to a direct generalization of the standard theory.  For this specific type of observer, the existence of a maximal proper acceleration implies the existence of a maximal Unruh temperature. An upper bound profile for temperature depending on the bunch population may be possible to be tested experimentally in current or near future experiments in laser-plasma acceleration, since the scales of acceleration required are not far from the upper bound  of the form \eqref{valueofthemaximalacceleration2} discussed in this paper.

The existence of a relation between maximal proper acceleration and maximal temperature has been discussed previously in different frameworks associated with quantum mechanical arguments \cite{CaianielloLandi,Brandt1983,Parentani Potting}.
In contrast, the theory of spacetimes of maximal proper acceleration considered in this paper is a classical framework. In fact we have considered the possibility and consequences of maximal proper acceleration in electrodynamics. In fact, for theory discussed here, the scale of the maximal proper acceleration involved depends on the specie of particle being accelerated in a form that the value of the maximal proper acceleration drops with the size of the charge and mass of the particle or bunch being accelerated with $1/N$, where $N$ is the number of particles in a single bunch \cite{Ricardo 2019,Ricardo 2024}. Specifically, the maximal proper acceleration scale for the  bunches considered in our model is of order $10^{22} \,m/s^2$ for bunches with $N\sim 10^{8}$ particles. Such scale of maximal proper acceleration and the luminosity of the required bunches are not far from current acceleration scales of order $10^{22}\,m/s^{2}$ for populations of order $10^8$ reached in experiments \cite{Wang et al. 2013,Kurz et al. 2021}, suggesting the realistic possibility of probing the spacetime structure by looking at the Unruh temperature effects in laser-plasma acceleration experiments. We have shown that the Unruh temperature formula can be extended to the case of hyperbolic observers in spacetimes of maximal proper acceleration.

The upper bound in proper acceleration implies an upper bound on the Unruh temperature, a bound that could depend on the system. For the bunches of electrons in laser-plasma acceleration considered, the maximal Unruh temperature is of order $10^2\,K$. The exploration of Unruh radiation to test the spacetime structure by means of laser-plasma acceleration was already suggested by Chen and Tajima \cite{Chen Tajima 1999}, who thoroughly investigated the observability of the effect with respect to other forms of radiations, most notably Larmor radiation. In our theory, Larmor's law is equivalent to the standard one \cite{Ricardo 2017}. Because of this, the result discussed here goes beyond the studies in \cite{Chen Tajima 1999} by showing that the existence of a maximal proper acceleration can be tested by looking for experimental bounds on the Unruh temperature profile in laser-plasma acceleration. For the model discussed in this paper, the upper bound on the acceleration decreases with the population $N$ by a factor of the form $1/N$ and on the species of particle.

 At first glance, one could expect that the value of the maximal acceleration for our model, given by $A_{\textrm{max}}=\,\frac{1}{N}\,\frac{m_0}{q^2}$, will be affected by a {\it soccer-ball like problem}, as in quantum gravity phenomenology \cite{Hossenfelder}: as $N$ becomes large, the system will not be able to be accelerated, since $A_{\textrm{max}}$ goes to zero. However, in principle such a conclusion is unrealistic in our case, since the model of bunch as a sole particle is an effective model for situations of stable bunches. Stability of the bunch requires that the space charge effects are under control in the sense that they are small \cite{Ricardo 2024}. These limitations on the domain of applicability of the model precludes the soccer-ball type conclusions and more generally, the applicability of the theory to large charged systems. Also, let us remark that our theory is not disconfirmed by the temperature scales reached inferred from the analysis of from 1D Planck-like spectra in beta decay processes \cite{Lynch}, with scales of order $10^{10}\, K$.  Clearly, in such phenomena, apart from electrodynamics we found the weak interaction involved. Moreover, our prediction of a temperature of order $10^2 \, K$ refers to the Unruh temperature associated with the bunch of particles as a coherent dynamical system, while for a sole electron, the maximal Unruh temperature associated with electromagnetism is $T_{\textrm{max}}(e)\sim \,10^{10} \,K$.

Let us discuss certain common points of our theory with other approaches to the same or related topics.
While in the theories of maximal proper acceleration of Brandt and of Caianiello and Landi \cite{Brandt1983,CaianielloLandi} the Unruh temperature formula was assumed as valid, in ref. \cite{Benedetto Feoli 2015} a derivation of it in the context of Caianiello's maximal acceleration theory was carried out, with the conclusion that the existence of a maximal proper acceleration does not imply a bound on the Unruh temperature, in apparent contradiction with the assumptions taken by Brandt and by Caianiello and Landi and with the conclusions of the present research. Besides referring to  different theories than ours, we think that the discrepancy in the reasoning and conclusions in \cite{Benedetto Feoli 2015} and ours originates in the methodology used. In \cite{Benedetto Feoli 2015} the authors started their discussion assuming Minkowski spacetime and by writing the standard relation between Cartesian coordinates and Rindler coordinates for the case of an hyperbolic observer in Minkowski space (equations (3) and (4) in ref. \cite{Benedetto Feoli 2015}). After that, the authors re-casted the relation in terms of the proper time of the metric of maximal acceleration and carried out a parallel analysis to the analysis of Alsing and Milonni of the standard (1+1)-dimensional case \cite{Alsing Milonni}. Our point of disagreement is that for us the physical proper time in our theory and hence, the one that should be used in the definition of the Rindler coordinates, is the proper time $s[\vartheta]$ calculated with the metric of maximal proper acceleration, given by the expression \eqref{propertime g}, and not the one calculated by means of the Minkowski metric as Benedetto and Feoli did. Since for hyperbolic observers, the proper time $s[\vartheta]$ calculated using $g$ and the proper time $\tau[\vartheta]$ are related by a constant factor $(1-a^2/A^2_{\textrm{max}})$, there is a discrepancy in the definition of the Rindler coordinates between our suggested methodology and the one used in \cite{Benedetto Feoli 2015}, a difference in definition that accounts for the difference in the Unruh temperature formula given by the two theories.

The analysis developed in this paper can be adapted to other theories with a maximal proper acceleration. However, for the theories where the dependence in $(m,q)$ is homogeneous of degree zero in the ratio $q/m$, as in Maxwell-Lorentz linear electrodynamics and in Born-Infeld non-linear electrodynamic theory or it has a dependence such that the associated maximal proper acceleration grows linearly with the mass of the specie of particle, like in Caianiello' theory, then the phenomenological consequences of maximal proper acceleration in the way discussed in this paper are less apparent to observe in laser-plasma experiments. There is also the additional problem of motivating the generalization of the Unruh temperature formula in such schemes. For instance, let us consider Caldirola's theory of discrete electrodynamics. In that theory, the existence of a minimum unit of time leads directly to a maximal acceleration that depends upon the ratio $m/q^2$ \cite{Caldirola}, which is the same expression for maximal acceleration \eqref{valueofthemaximalacceleration} except for a numerical factor of order $1$. The difficulty in extending our analysis to Caldirola's theory resides in the difficulty of establishing the Unruh temperature formula for such theory. This requires the construction of several notions like the energy-momentum four momentum and an adequate dispersion relation. Being Caldirola's theory discrete in nature, it is not clear how such requirements can be achieved properly. This is an interesting problem to pay attention.

Let us also remark on the classical theory developed by M. Goto et al. \cite{Goto et al. 2010}, where it was argued that the strong equivalence principle holds good up to values of the gravitational field acceleration in our electrodynamic theory. The justification of the spaces of maximal proper acceleration proposed in the present paper resides partially in a generalized form of Einstein's equivalence principle. It will be very interesting to understand in a deeper form the relation between both arguments.

Merit to mention is that our result showing the dependence of the rest mass with acceleration as given by the relation \eqref{modified rest mass} can be seen as an analogous result to the one found in \cite{Castro-Perelman 2024}, where the mass parameter depends on the temperature. In both theories, such a relation is established by means of the Unruh temperature formula, whose validity has been argued in this paper for spacetimes of maximal proper acceleration. However, in our theory, the  general expression for the rest mass measured by an observer \eqref{mass measured by O}, the Unruh temperature formula \eqref{temperature} and its link with the mass by means of the relation \eqref{mass in term of temperature} are not linked to the Planck scale, as it is the case Castro-Perelman theory \cite{Castro-Perelman 2024}. This could serve eventually to discriminate between both approaches.

Finally, let us remark again that our theory is a classical theory. Several of the approaches where a maximal proper acceleration makes its appearance  are quantum mechanical models. This is a surprising consequence, since by the Copenhagen interpretation of quantum mechanics, there is no classical trajectories and hence, there is no meaning of a classical notion of acceleration. However, if we take seriously such notions of maximal proper acceleration as Caianiello's quantum mechanical derivation or Schwinger's maximal acceleration discussed above, then a different interpretation of quantum evolution than the orthodox interpretation is necessary. One clear option is to adopt the de Broglie-Bohm interpretation \cite{Bohm}. Another possibility is to consider Nelson stochastic quantum mechanics \cite{Nelson}, although some level of modification in the definition of acceleration should be considered in this case. The point of view supported by the author of this paper supports an emergent view of quantum mechanics, where classical trajectories emerge, although they have a fluctuating character \cite{Ricardo05b, Ricardo06, Ricardo2014, Ricardo 2024b}. It is very exciting that the idea of maximal proper acceleration can be used to probe such a fundamental problems: if a classical notion of acceleration can be probed valid at scales where pure quantum fluctuations are expected to delete any sign of classicality, it can confirm the need of an alternative view to quantum effects than the proposed by orthodox interpretations.

\small{
}

\end{document}